\documentclass[aps,prd,showpacs,showkeys,preprintnumbers]{revtex4}
\usepackage{graphicx}                              
\usepackage{amsmath,amsfonts}
\graphicspath{{./pictures/}}                       

\usepackage{epsfig}
\usepackage{amssymb}
\usepackage{amsfonts}
\usepackage{float}
\usepackage{bbm}
\usepackage{psfrag}
\newcommand{\bdm}{\begin{displaymath}}
\newcommand{\edm}{\end{displaymath}}
\newcommand{\beq}{\begin{equation}}
\newcommand{\eeq}{\end{equation}}
\newcommand{\bea}{\begin{eqnarray}}
\newcommand{\eea}{\end{eqnarray}}
\newcommand{\bit}{\begin{itemize}}
\newcommand{\eit}{\end{itemize}}
\newcommand{\bc}{\begin{center}}
\newcommand{\ec}{\end{center}}
\newcommand{\re}{\relax{\rm I\kern-.18em R}}
\newcommand{\ID}{\mathbbm{1}}

\newcommand{\N}{\mathbbm{N}}

\newcommand{\Comp}{\mathbbm{C}}

\newcommand{\ie}{{\it i.e. }}
\newcommand{\D}{{\cal D}^{(ov)}}

\newcommand{\SD}{\hat{\cal D}^{(ov)}}

\newcommand{\SDw}{\hat{\cal D}^{(W)}}

\newcommand{\ImpSpace}{{\cal P}}

\newcommand{\sumFL}{\sum\limits_{i=1}^{N_f}}

\newcommand{\fermiMat}{{\cal M}}
\newcommand{\kapCrit}{\kappa_\mathrm{crit}}
\newcommand{\mAvg}{{\langle m \rangle}}
\newcommand{\sAvg}{{\langle s \rangle}}

\begin{document}
\preprint{HU-EP-07/28, DESY 07-108}

\title{The phase structure of a chirally invariant lattice Higgs-Yukawa model - \\
numerical simulations} 

\author{P. Gerhold$^a$, K. Jansen$^b$}
\affiliation{$^a$Humboldt-Universit\"at zu Berlin, Institut f\"ur Physik, 
Newtonstr. 15, D-12489 Berlin, Germany\\
$^b$DESY,\\
 Platanenallee 6, D-15738 Zeuthen, Germany}

\date{July 25, 2007}

\begin{abstract}
The phase diagram of a chirally invariant lattice Higgs-Yukawa model 
is explored by means of numerical simulations. The results revealing 
a rich phase structure are compared to analytical large $N_f$ 
calculations which we performed earlier. 
The analytical and numerical results are in excellent agreement at large
values of $N_f$. In the opposite case the large $N_f$ computation
still gives a good qualitative description of the phase diagram. In particular
we find numerical evidence for the predicted ferrimagnetic phase at intermediate
values of the Yukawa coupling constant and for the symmetric phase at strong 
Yukawa couplings. 
Emphasis is put on the finite size effects which can hide the existence
of the latter symmetric phase.
\end{abstract}



\maketitle

\section{Introduction}
\label{sec:Introduction}

The main target of lattice studies of the Higgs-Yukawa sector of the 
electroweak standard model is the non-perturbative determination of lower 
and upper bounds of the Higgs boson mass~\cite{Holland:2003jr,Holland:2004sd} 
as well as its decay properties. There are two main developments which 
warrant to reconsider these questions: first, with the advent of the LHC, 
we are to expect that properties of the standard model Higgs boson, such as 
the mass and the decay width, will be revealed experimentally. Second, there 
is, in contrast to the situation of earlier investigations of lattice 
Higgs-Yukawa models, a consistent formulation of an exact lattice chiral 
symmetry~\cite{Luscher:1998pq} based on the Ginsparg-Wilson 
relation~\cite{Ginsparg:1981bj}. 

Before questions of the Higgs mass bounds and decay properties can be addressed, 
the phase structure of the model needs to be investigated in order to determine 
the (bare) couplings in parameter space where eventual simulations of 
phenomenological interest can be performed. There has been a large activity of 
investigating lattice Higgs-Yukawa models in the past, see e.g. 
Refs.~\cite{Smit:1989tz,Shigemitsu:1991tc,Golterman:1990nx,book:Montvay,book:Jersak,Golterman:1992ye,Jansen:1994ym} 
for reviews. In particular, the phase structure of lattice Higgs-Yukawa models
was investigated in great detail, see e.g. 
Refs.~\cite{Hasenfratz:1989jr,Lee:1989mi,Bock:1990tv,Lin:1991cs,Hasenfratz:1991it,Hasenfratz:1992xs,Bock:1992yr,Bock:1997fu,Bock:1999qa} 
for a still incomplete list. However, in these investigations, the lattice 
formulation of the corresponding Higgs-Yukawa theory broke explicitly chiral symmetry. 

This situation changed when it was realized that the Ginsparg-Wilson 
relation~\cite{Ginsparg:1981bj} leads to the notion of an exact lattice chiral 
symmetry~\cite{Luscher:1998pq} allowing thus to go beyond 
the earlier models. Based on this development, 
the interest in lattice studies of Higgs-Yukawa models has been 
renewed~\cite{Bhattacharya:2006dc,Giedt:2007qg,Poppitz:2007tu,KutiPrivateCom,Gerhold:2007yb}.
Here, we follow L\"uscher's proposition for a chirally invariant and hence consistent 
lattice Higgs-Yukawa model given in Ref.~\cite{Luscher:1998pq}. 
In particular, we want to address here the question whether the phase 
structure remains as complex as has been found out in the earlier work mentioned above. 
Additional and phenomenologically more interesting questions concerning  
the behaviour of the renormalized Higgs and Yukawa couplings will be 
addressed in future works.

In \cite{Gerhold:2007yb} we have studied the phase structure of this chirally invariant 
Higgs-Yukawa model by means of a large $N_f$ computation, where $N_f$ denotes the number 
of fermion generations. We found a complex phase structure that resembles qualitatively 
the one of earlier lattice Higgs-Yukawa models, see e.g.~\cite{Bock:1990tv}, which, however, 
were lacking chiral symmetry. In the present paper we want to confront the results 
obtained in the large $N_f$ approximation with direct numerical simulations for finite 
values of $N_f$. 
We remark that in the present work, as in Ref.~\cite{Gerhold:2007yb}, 
we neglect the gauge degrees of freedom and consider the pure scalar-fermion sector of the 
electroweak standard model.

To be more specific, we consider here a four-dimensional, chirally invariant 
$SU(2)_L\times SU(2)_R$ Higgs-Yukawa model discretized on a finite lattice with $L$ 
lattice sites per dimension such that 
the total volume becomes $V=L^4$. We set the lattice spacing to one throughout the paper.
The model contains one four-component, real Higgs field $\Phi$ and we consider $N_f$ fermion 
doublets represented by eight-component spinors $\psi^{(i)}$, $\bar\psi^{(i)}$ with $i=1,...,N_f$. 
Furthermore, there are $N_f$ {\it auxiliary} fermionic doublets $\chi^{(i)}$, $\bar\chi^{(i)}$ 
serving as a construction tool in the creation of a chirally invariant Yukawa interaction 
term. Once the chiral invariance is established these {\it unphysical} fields are 
integrated out leading to a more complex model depending only on the Higgs field $\Phi$ and 
the $N_f$ {\it physical} fermion doublets $\psi^{(i)}$. The partition function is  written as 
\beq
Z = \int D\Phi\,\prod\limits_{i=1}^{N_f} \left[D\psi^{(i)}\, D\bar\psi^{(i)}\, D\chi^{(i)}\,
D\bar\chi^{(i)} \right]\,  \exp\left( -S_\Phi -S_F^{kin} - S_Y  \right)
\eeq
with the total action being decomposed into the Higgs action 
$S_\Phi$, the kinetic fermion action $S_F^{kin}$, 
and the Yukawa coupling term $S_Y$. It should be stressed once 
again that {\it no gauge fields} are included within this model. 

The kinetic fermion action describes the propagation of the 
physical fermion fields $\psi^{(i)}$,$\bar\psi^{(i)}$ 
in the usual manner according to 
\beq
S_F^{kin} = \sumFL \sum\limits_{n,m} \bar\psi^{(i)}_n \D_{n,m} \psi^{(i)}_{m} -
2\rho\bar\chi^{(i)}_n\ID_{n,m} \chi^{(i)}_m
\eeq
where the four-dimensional coordinates $n,m$ as well as all field 
variables and coupling constants are given in lattice units
throughout this paper. The (doublet) Dirac operator $\D= \SD\otimes\SD$ 
is given by the Neuberger overlap operator $\SD$, which is related to the Wilson operator 
$\SDw=\gamma^E_\mu \frac{1}{2}(\nabla_\mu^f + \nabla_\mu^b) - \frac{r}{2} \nabla^b_\mu\nabla^f_\mu$ by 
\bea
\SD &=& \rho\left\{1+\frac{\hat A}{\sqrt{\hat A^\dagger \hat A}}   \right\},\quad \hat A = \SDw - \rho, \quad 1\le \rho < 2r
\eea
with $\nabla^f_\mu$, $\nabla^b_\mu$ denoting the forward and backward difference quotients.
Note that in absence of gauge fields this kinetic part  corresponds 
to the one of free fermions which will be exploited in the 
numerical construction of the overlap operator later.
In particular, the eigenvalues of $\SD$ can be computed 
analytically. In momentum space with the allowed four-component momenta
\beq
p \in \ImpSpace = \Bigg\{
\begin{array}{*{3}{c}}
(-\pi,\pi]^{\otimes 4} & : & \mbox{for } L = \infty \\
\{2\pi n/L\,:\, n\in \N_0, n<L\}^{\otimes 4} & : & \mbox{for } L < \infty \\
\end{array}
\eeq
the eigenvalues of the doublet operator $\D$ are given by
\bea
\label{eq:EigenValuesOfD}
\nu^\epsilon(p)&=& \rho + \rho\cdot\frac{\epsilon i\sqrt{\tilde p^2} + 2r\hat p^2 - \rho}{\sqrt{\tilde p^2 + (2r\hat p^2 -
\rho)^2}},\quad \tilde p_\mu = \sin(p_\mu),\quad \hat p_\mu = \sin\left(\frac{p_\mu}{2}\right), \quad \epsilon = \pm 1.
\eea

We remark that the auxiliary fields $\chi^{(i)}$ do not propagate at all 
and that their contribution to $S_F^{kin}$ is chosen such that the model 
will obey an exact lattice chiral symmetry.

The Higgs field couples to the fermions according to the Yukawa coupling term
\beq
S_Y = y_N \sum\limits_{n,m}\sumFL(\bar\psi^{(i)}_n+\bar\chi^{(i)}_n) \underbrace{\left[
\ID_{n,m}\frac{(1-\gamma_5)}{2}\phi_n 
+ \ID_{n,m}\frac{(1+\gamma_5)}{2}\phi^{\dagger}_n  \right]}_{B_{n,m}} (\psi^{(i)}_m+\chi^{(i)}_m)
\label{eq:DefYukawaCouplingTerm}
\eeq
where $y_N$ denotes the Yukawa coupling constant and $B_{n,m}$ 
will be referred to as Yukawa coupling matrix. 
Here the Higgs field $\Phi_n$ is rewritten as a 
quaternionic, $2 \times 2$ matrix 
$\phi_n = \Phi_n^0\ID -i\Phi_n^j\tau_j$, with $\vec\tau$
denoting the vector of Pauli matrices,
acting on the $SU(2)$ index of the fermionic doublets.
Due to the chiral character of this model, 
left- and right-handed fermions couple differently to the 
Higgs field, as can be seen from the appearance of the 
projectors $(1\pm \gamma_5)/2$ in the Yukawa term.
Multiplying out the involved Gamma- and Pauli-matrices one can 
rewrite the coupling matrix in the compactified form
\bea
B_{m,n} = \delta_{m,n} \cdot \hat B(\Phi_n), \quad
\hat B(\Phi_n) =  
\left(
\begin{array}{*{2}{c}}
\Phi_n^0\ID+i\Phi_n^3\gamma_5 & \Phi_n^2\gamma_5 +i\Phi_n^1\gamma_5  \\
-\Phi_n^2\gamma_5 +i\Phi_n^1\gamma_5 & \Phi_n^0\ID-i\Phi_n^3\gamma_5  \\
\end{array}
\right)
\label{eq:DefBHat}
\eea
being block diagonal in position space.
The model then obeys an exact, but lattice modified, chiral 
symmetry according to
\bea
\delta\psi^{(i)}=i\epsilon\left[\gamma_5\left(1-\frac{1}{2\rho}\D \right)\psi^{(i)} + \gamma_5\chi^{(i)}\right], & 
\delta\chi^{(i)} = i\epsilon\gamma_5\frac{1}{2\rho}\D\psi^{(i)},& 
\delta\phi =2i\epsilon\phi    \\
\delta\bar\psi^{(i)}=i\epsilon\left[\bar\psi^{(i)}\left(1-\frac{1}{2\rho}\D \right)\gamma_5 + \bar\chi^{(i)}\gamma_5\right], & 
\delta\bar\chi^{(i)} = i\epsilon\bar\psi^{(i)}\frac{1}{2\rho}\D\gamma_5,& 
\delta\phi^\dagger =-2i\epsilon\phi^\dagger    
\eea
recovering the chiral symmetry in the continuum 
limit~\cite{Luscher:1998pq}.

Finally, the lattice Higgs action is given by the usual lattice notation
\beq
\label{eq:LatticePhiAction}
S_\Phi = -\kappa_N\sum_{n,\mu} \Phi_n^{\dagger} \left[\Phi_{n+\hat\mu} + \Phi_{n-\hat\mu}\right]
+ \sum_{n} \Phi^{\dagger}_n\Phi_n + \lambda_N \sum_{n} \left(\Phi^{\dagger}_n\Phi_n - N_f \right)^2
\eeq
with the only particularity that the fermion generation number $N_f$ appears in the quartic
coupling term which was a convenient convention for the large $N_f$ analysis. However, this
version of the lattice Higgs action is equivalent to the usual continuum notation
\bea
\label{eq:ContinuumPhiAction}
S_\varphi &=& \sum_{n} \left\{\frac{1}{2}\left(\nabla^f_\mu\varphi\right)_n^{\dagger} \nabla^f_\mu\varphi_n
+ \frac{1}{2}m_0^2\varphi_n^{\dagger}\varphi_n + \lambda_0\left(\varphi_n^{\dagger}\varphi_n\right)^2   \right\},
\eea
with the bare mass $m_0$ and the bare quartic coupling constant $\lambda_0$. The connection is
established through a rescaling of the Higgs field and the involved coupling constants according
to
\beq
\label{eq:RelationBetweenHiggsActions}
\varphi_n = \sqrt{2\kappa_N}\Phi_n, \quad
\lambda_0=\frac{\lambda_N}{4\kappa_N^2}, \quad
m_0^2 = \frac{1 - 2N_f\lambda_N-8\kappa_N}{\kappa_N}, \quad
y_0 = \frac{y_N}{\sqrt{2\kappa_N}}
\eeq
where $y_0$ denotes the Yukawa coupling constant corresponding to the continuum
notation.

\section{Simulation Algorithm}
\label{sec:SimAlgorithm}

The first step towards a numerical treatment of the considered Higgs-Yukawa model is
to integrate out the fermionic degrees of freedom leading to the
effective action
\beq
\label{eq:effectiveHiggsAction1}
S_{eff}[\Phi]= S_\Phi[\Phi] - N_f\cdot \log\det\left(\fermiMat\right)
\eeq
where the fermionic matrix
\beq
\label{eq:DefOfFermiMat}
\fermiMat = y_NB\D -2\rho\D -2\rho y_N B 
\eeq
was given in Ref.~\cite{Gerhold:2007yb}. 

Since we focus here on checking the validity of our earlier analytical investigation of the 
phase structure, which was determined in the large $N_f$-limit, we 
will only consider even values for $N_f$, 
allowing to rewrite the effective action according to 
\beq
\label{eq:effectiveHiggsAction2}
S_{eff}[\Phi]= S_\Phi[\Phi] - \frac{N_f}{2}\cdot \log\det\left(\fermiMat\fermiMat^\dagger\right)
\; , \;\;\; N_f\;\;\; \mathrm{even}\; .
\eeq
Thus the positivity of the determinant in Eq.~(\ref{eq:effectiveHiggsAction2}) is guaranteed. 

For the numerical treatment of the remaining determinant in Eq.~(\ref{eq:effectiveHiggsAction2}) 
we have implemented an Hybrid-Monte-Carlo (HMC) algorithm~\cite{Duane:1987de,Gottlieb:1987mq},
with $N_f/2$ complex pseudo-fermionic fields $\omega_j$ according to the HMC-Hamiltonian
\beq
H(\Phi,\xi,\omega_j) = S_\Phi[\Phi] + \frac{1}{2}\xi^\dagger \xi 
+ \sum\limits_{j=1}^{N_f/2}\frac{1}{2}\omega_j^\dagger \left[\fermiMat\fermiMat^\dagger\right]^{-1}\omega_j 
\eeq
where $\xi$ denote the real momenta, conjugate to the Higgs field $\Phi$. 

The application of the matrix $[\fermiMat\fermiMat^\dagger]^{-1}$ on $\omega_j$ can then
be performed by means of a Conjugate Gradient algorithm due to the hermiticity of
$\fermiMat\fermiMat^\dagger$. However, for the computation of $\fermiMat x$, where $x$ is an 
arbitrary vector, we exploit the fact that there are no gauge
fields included within our model. The eigenvectors of the used Neuberger overlap operator
$\D$ are therefore explicitly known to be the plane waves
\beq
\Psi_n^{p,\zeta\epsilon k} = e^{i p \cdot n} \cdot u^{\zeta\epsilon k}(p),\quad
u^{\zeta\epsilon k}(p) = \sqrt{\frac{1}{2}}
\left(
\begin{array}{*{1}{c}}
u^{\epsilon k}(p) \\
\zeta u^{\epsilon k}(p) \\
\end{array}
\right), \quad
\zeta=\pm 1, \,
\epsilon=\pm 1, \,
k \in \{1,2\}
\eeq
with $u^{\epsilon k}(p)$ denoting the usual four-component spinor structure 
\beq
u^{\epsilon k}(p) = 
\sqrt{\frac{1}{2}}
\left(
\begin{array}{*{1}{c}}
\xi_k \\
\epsilon\frac{\tilde p \bar\Theta}{\sqrt{\tilde p^2}} \xi_k 
\end{array}
\right)
\,
\mbox{for }\tilde p\neq 0
\quad
\mbox{and}
\quad
u^{\epsilon k}(p) = 
\sqrt{\frac{1}{2}}
\left(
\begin{array}{*{1}{c}}
\xi_k \\
\epsilon \xi_k \\
\end{array}
\right)
\,
\mbox{for }\tilde p= 0.
\eeq
Here $\xi_k\in\Comp^2$ are two orthonormal vectors and the four component 
quaternionic vector $\bar\Theta$
is defined as $\bar\Theta = (\ID, i\vec\tau)$.
The corresponding eigenvalues $\nu^\epsilon(p)$ were given in Eq.~(\ref{eq:EigenValuesOfD}).
The operators $B$ and $\D$ are thus both block-diagonal, the first in position space and the
latter in momentum space. In our approach we use a Fast Fourier Transform (FFT)~\cite{FFTW05} to switch between the position and momentum representations, such
that all operator applications can be trivially performed due to their actual
block-diagonal structure. This is particularly advantageous for the overlap operator,
since the usual construction of this operator would be based on very demanding approximations, e.g. polynomial approximations.

A second advantage of this approach is that the applied Dirac operator can easily be replaced by other
operators simply by adopting the corresponding eigenvalues. 

Concerning the parallelization of the program there are several options. For example there
are efficient parallelized FFT-routines available~\cite{FFTW05}. Here, however, we use a
trivial - but very efficient - parallelization which is possible due to the large number
of fermion generations $N_f$. We simply perform each of the $N_f/2$ force calculations on
a separate computer node. 

For the integration of the obtained forces we find the Leap-Frog integration scheme to be 
efficient on small lattices. This situation changes with increasing lattice size and for $L\ge 16$
we get better performance with higher order integrators. In that case we use an order 4 
Omelyan-integrator~\cite{Omelyan:2003,Takaishi:2005tz}. The integration is then 
performed over a fixed
trajectory length set to unity with the typical value $\epsilon=0.1$ for the step size. 
The step size $\epsilon$ is chosen such that the acceptance rate stays between $80\%$ and $95\%$. 

The observables we will be using for exploring the phase structure 
are the {\it magnetization} $m$ and the {\it staggered magnetization} $s$, 
\begin{equation} 
m = \left[\sum\limits_{i=0}^3\Big|\frac{1}{L^4}\sum\limits_{n} \Phi_n^i\Big|^2\right]^{\frac{1}{2}}
\label{magnetizations} , \quad
s = \left[\sum\limits_{i=0}^3\Big|\frac{1}{L^4}\sum\limits_{n} \left(-1\right)^{\sum\limits_\mu n_\mu}\cdot\Phi_n^i\Big|^2\right]^{\frac{1}{2}}
\end{equation}
and the corresponding susceptibilities
\begin{equation} 
\chi_m =  V\cdot \left[\langle m^2\rangle -\langle m \rangle^2 \right], 
\quad
\chi_s =  V\cdot \left[\langle s^2\rangle -\langle s \rangle^2 \right],
\label{magneticsus}
\end{equation} 
where $\langle ... \rangle$ denotes the average over the $\Phi$-field 
configurations generated in the
Monte-Carlo process.

\begin{figure}[htb]
\begin{tabular}{cc}
\includegraphics[width=0.32\textwidth]{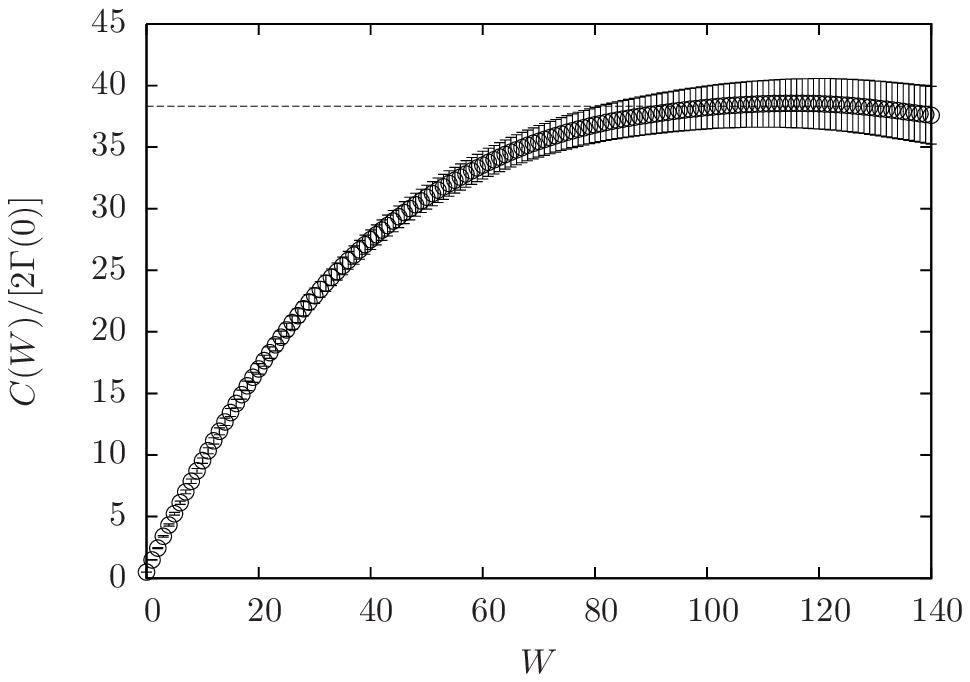}&
\includegraphics[width=0.32\textwidth]{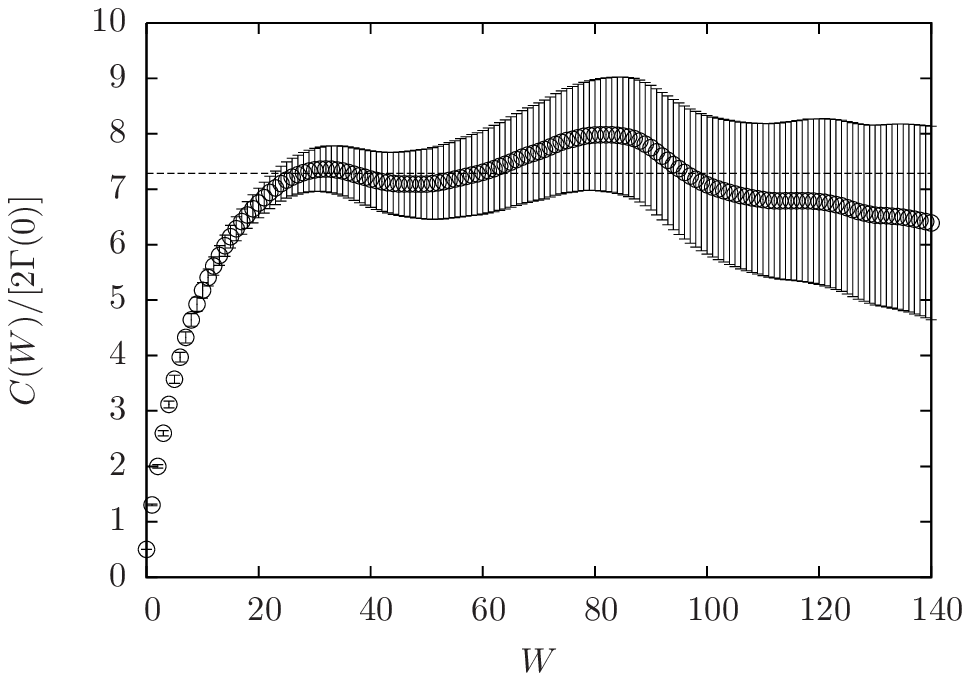}
\\
(a) & (b) \\
\end{tabular}
\caption{Example for the determination of the auto-correlation time for two 
different points in the phase 
diagram at $\lambda_N=0.05$, $L=16$, $N_f=2$, and $y_N=30$. 
(a): A point very close to the phase transition with $\kappa_N=0.042$. 
(b): A point farer away from the phase transition line with $\kappa_N=0.060$.}
\label{fig:DetOfCorTime}
\end{figure}

The auto-correlation of our measurements of these observables in the 
Monte Carlo time $t$ is then accounted for by applying the 
$\Gamma$-strategy~\cite{Wolff:2003sm}. 
In this approach the error $\sigma_A$ of an observable $A$ is
rewritten as a sum over the correlation function $\Gamma(t)$ according to
\beq
\label{eq:DefGammaStrat}
\sigma_A^2=\frac{C(\infty)}{N},\quad
C(W) = \sum\limits_{t=-W}^W \Gamma(t),\quad
\Gamma(t) = \frac{1}{N-|t|} \sum\limits_i \left[ A^{(i)}-\langle A\rangle  \right] \cdot 
\left[ A^{(i+t)}-\langle A\rangle  \right]
\eeq
where $A^{(i)}$ denotes the measurement of the observable $A$ in the $i$-th configuration
and $N$ is the total number of collected configurations.
The variable $W$ is the window in which the function $\Gamma(t)$ is to be summed up.
It should be large enough to obtain reliable estimates of the auto-correlation time
$\tau$ which is defined through the exponential decay 
rate of $\Gamma(t)$
\beq
\Gamma(t) \propto \exp\left({-\frac{|t|}{\tau}}\right)
\eeq
and is thus directly connected to the sum $C(W)$ of the auto-correlation 
function $\Gamma(t)$. Typical 
examples for the determination of $C(\infty)$ by fitting the function $C(W)$ to a constant are 
presented in Fig.~\ref{fig:DetOfCorTime}. 
Since the auto-correlation length $\tau$ depends strongly on the distance to the 
phase transition
we have selected one point in the parameter space close to the phase transition 
(Fig.~\ref{fig:DetOfCorTime}a) and one point farer away from it 
(Fig.~\ref{fig:DetOfCorTime}b). Both points correspond, however,
to the ferromagnetic phase with a non-vanishing Higgs field expectation value, 
\ie $\mAvg>0$. 
For the Higgs field magnetization $m$ as the underlying observable 
we find in these examples the auto-correlation times 
$\tau_m^{(a)}=38.3\pm 1.9$ in Fig.~\ref{fig:DetOfCorTime}a and
$\tau_m^{(b)}=7.3\pm 0.4$ in Fig.~\ref{fig:DetOfCorTime}b.
We remark here that the value obtained in Fig.~\ref{fig:DetOfCorTime}a
is the largest auto-correlation time for the magnetization $m$
encountered in our studies.
Although the auto-correlation time indeed increases when approaching the 
phase transition, its value remains acceptable for our purposes. Given that our 
typical statistics is $O(10^4)$ $\Phi$-field configurations, this leads to 
reliable error determinations for the physical quantities of interest.

\bc
\ec

\section{Results for small values of the Yukawa coupling constant}
\label{sec:SmallY}

In this section we will discuss the structure of the phase diagram at 
small values of the Yukawa coupling constant. All numerical results 
from our simulations are obtained by employing the algorithm as detailed 
in Section~\ref{sec:SimAlgorithm}. 
The anticipated structure of the phase diagram 
can be inferred from our large $N_f$ computation in Ref.~\cite{Gerhold:2007yb}. 
In that large $N_f$ approach the Higgs field and the coupling constants are
scaled according to
\beq
\label{eq:LargeNBehaviourOfCouplings1}
y_N = \frac{\tilde y_N}{\sqrt{N_f}}\,,\quad  
\lambda_N = \frac{\tilde \lambda_N}{N_f}\,,\quad
\kappa_N = \tilde \kappa_N\,,\quad
\Phi_n = \sqrt{N_f} \cdot\tilde\Phi_n\,,
\eeq
where the quantities $\tilde y_N$, $\tilde \kappa_N$, $\tilde \lambda_N$, and $\tilde\Phi_n$
are held constant in the limit $N_f \rightarrow\infty$.
Here, we want to confront this predicted phase structure with the results of 
our numerical simulations.

At small values of the Yukawa coupling constant, there are two phase transitions when varying $\kappa_N$:
the first is a phase transition from a ferromagnetic (FM) phase, with $\sAvg =0$ and 
$\mAvg >0$, to the symmetric (SYM) phase with $\mAvg=\sAvg=0$.
The second corresponds to a phase transition from the 
symmetric phase to an anti-ferromagnetic (AFM) phase with
$\mAvg=0$ and $\sAvg>0$. These phase transitions are expected to 
be of second order. To locate the phase transition points,
we decided to fit the data for the susceptibilities  $\chi_m$, $\chi_s$
in Eq.~(\ref{magneticsus}) as a function of $\kappa_N$ 
according to the -- partly phenomenologically motivated -- ansatz
\begin{equation} 
\label{finitesizesus}
\chi_{m,s} = A_1^{m,s}\cdot\left(\frac{1}{L^{-2/\nu} + A_{2,3}^{m,s}(\kappa_N-\kapCrit^{m,s})^2}\right)^{\gamma/2},
\end{equation} 
where $A_1^{m,s}$, $A_{2,3}^{m,s}$, and $\kapCrit^{m,s}$ are the fitting parameters 
for the magnetic susceptibility and staggered susceptibility, respectively, 
and $\nu$, $\gamma$ denote the critical exponents of the $\Phi^4$-theory. 
Here $A_{2,3}^m$ ($A_{2,3}^s$) is actually meant to refer to
two parameters, namely $A_2^m$ ($A_2^s$) for $\kappa_N<\kapCrit^m$ ($\kappa_N<\kapCrit^s$) 
and $A_3^m$ ($A_3^s$) in the other case, such that 
the resulting curve is not necessarily symmetric. The phase transition point is then given at the 
value of $\kappa_N=\kapCrit^m$ ($\kappa_N=\kapCrit^s$) where the magnetic (staggered) susceptibility 
develops its maximum.
We remark that  the ansatz in Eq.~(\ref{finitesizesus}), although not being 
unique, provides a very good description of our numerically obtained data leading 
to a reliable determination of the critical hopping parameters $\kapCrit^{m,s}$.

\bc
\begin{figure}[htb]
\begin{tabular}{ccc}
\includegraphics[width=0.32\textwidth]{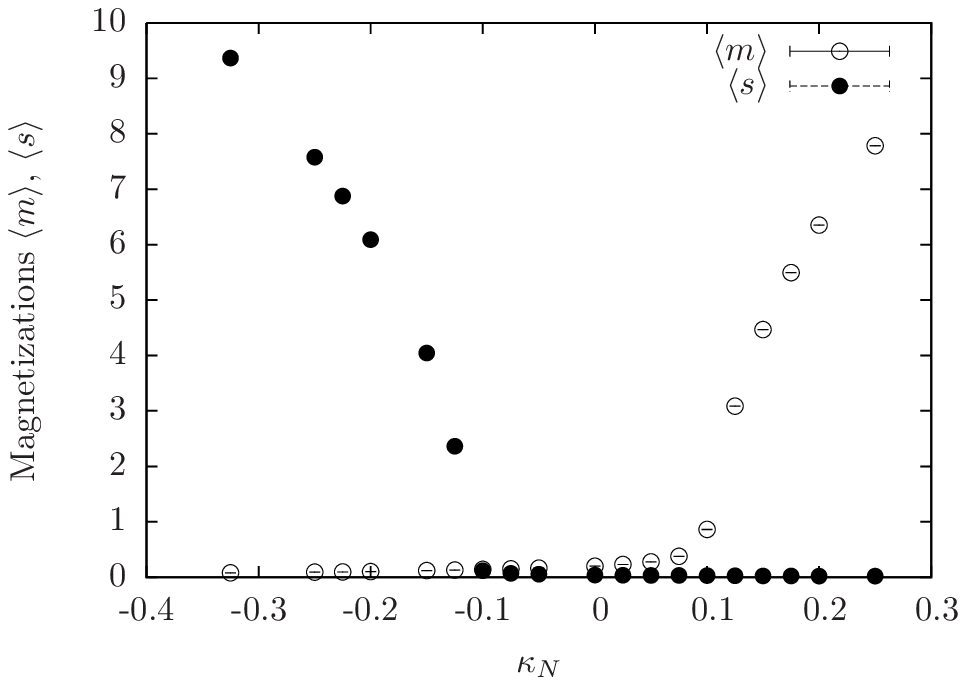}&
\includegraphics[width=0.32\textwidth]{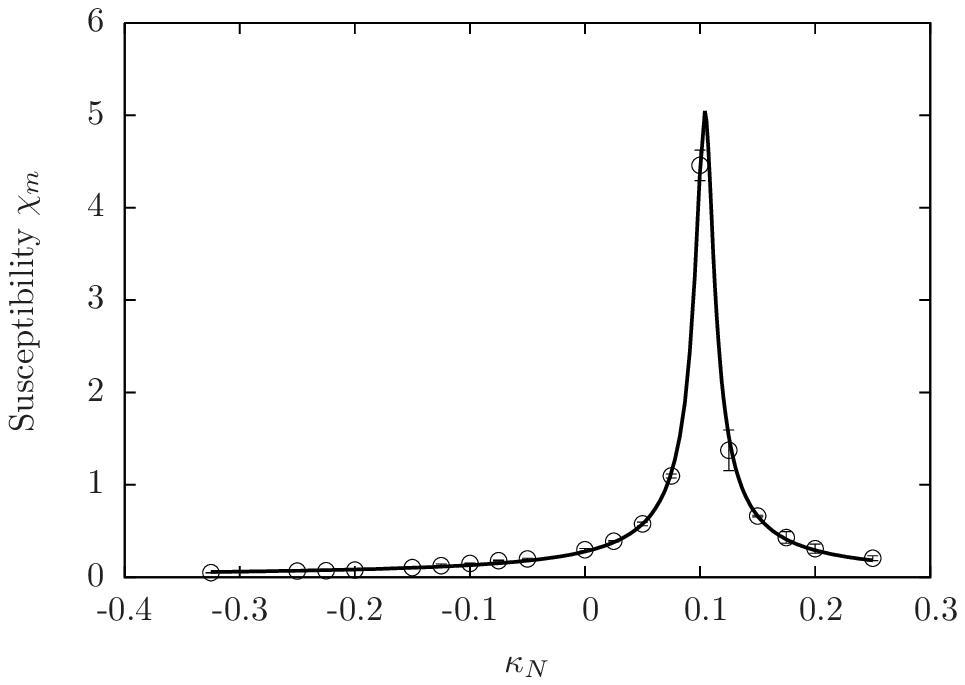}&
\includegraphics[width=0.32\textwidth]{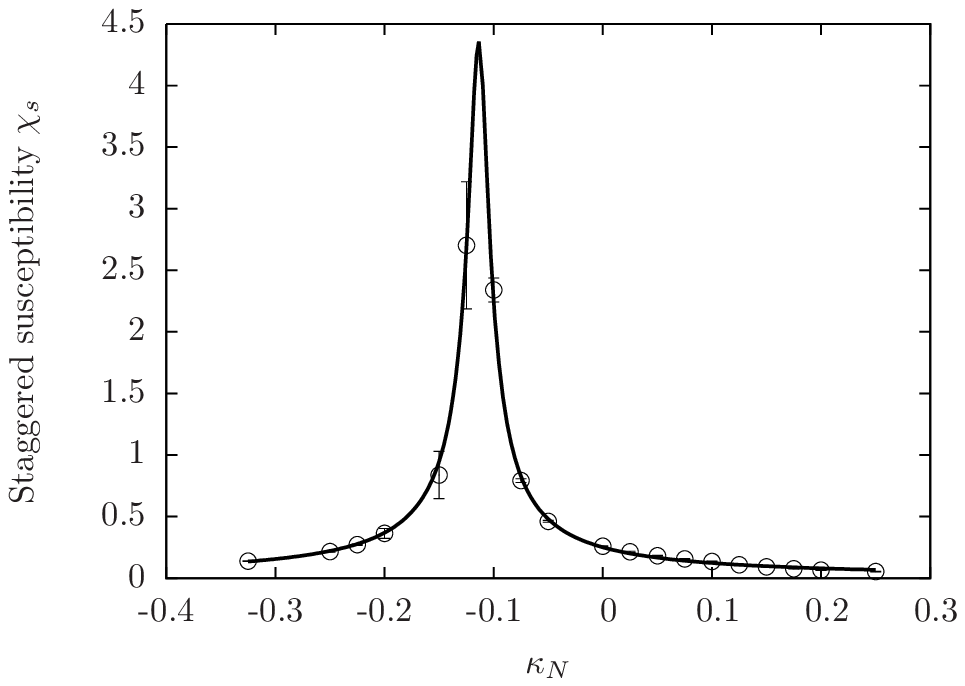}
\\
(a) & (b) & (c) \\
\end{tabular}
\caption{An example for the determination of the phase transition points 
separating the ferromagnetic and the anti-ferromagnetic phase from the 
symmetric phase. We show, as a function of $\kappa_N$ the behaviour of the 
average magnetization $\mAvg$ and staggered magnetization $\sAvg$ in panel (a). 
The corresponding susceptibilities are plotted in panels (b) and (c). 
The solid lines are fits to the finite size formula
of Eq.~(\ref{finitesizesus}). The parameters chosen are 
$\tilde y_N=0.632$, $\tilde\lambda_N=0.1$, $L=6$ and $N_f=10$.}  
\label{fig:kappascan1}
\end{figure}
\ec

In Fig.~\ref{fig:kappascan1} we present a typical example for the determination
of the phase transition points at small values of the Yukawa coupling constant.
The average magnetizations $\mAvg$ and $\sAvg$ as well as the corresponding
susceptibilities are shown as a function of $\kappa_N$. We clearly observe the 
vanishing of the magnetization and the staggered magnetization when the 
symmetric phase is entered (except for some small finite volume effects). 
Associated with these transitions are 
peaks in the susceptibilities. Note that the data for the susceptibilities 
are fitted very well using the ansatz of Eq.~(\ref{finitesizesus}), allowing 
for a good determination of the critical points. 

\bc
\begin{figure}[htb]
\includegraphics[width=5cm]{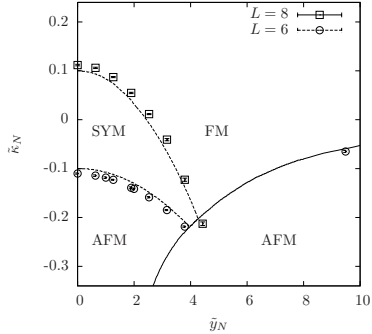}
\caption{The phase diagram at small Yukawa coupling constants together
with the $L=\infty$ prediction of the large $N_f$ calculation. 
The dashed lines denote second order phase transitions
while the solid line marks a first order transition.
The data with open squares were obtained on an $8^4$-lattice while the ones represented 
by open circles were measured on $6^4$-lattices. These results were obtained at
$\tilde \lambda_N=0.1$ and $N_f=10$.}
\label{fig:phasediagram1}
\end{figure}
\ec

Using the strategy just described we computed the values of $\kapCrit^m$ and $\kapCrit^s$ for various
Yukawa coupling constants $\tilde y_N<5$ holding the quartic coupling $\tilde\lambda_N=0.1$ 
constant. In Fig.~\ref{fig:phasediagram1} we summarize the numerical
results for the phase structure as obtained on $8^4$- and $6^4$-lattices at $N_f=10$ 
and compare them to the analytical $N_f=\infty$, $L=\infty$ phase structure. 
As expected we observe a symmetric (SYM), a ferromagnetic (FM) and an anti-ferromagnetic
(AFM) phase, with the symmetric phase bending strongly towards smaller values of the
critical hopping parameter when the Yukawa coupling constant is increased. 

As a general remark we note here that the simulations become extremely demanding when entering the
anti-ferromagnetic phase, due to an increasingly bad condition number of the fermionic matrix
$\fermiMat$. Within the anti-ferromagnetic phase we thus only present numerical results obtained on
$6^4$-lattices throughout this paper.

\bc
\begin{figure}[htb]
\begin{tabular}{cc}
\includegraphics[width=0.32\textwidth]{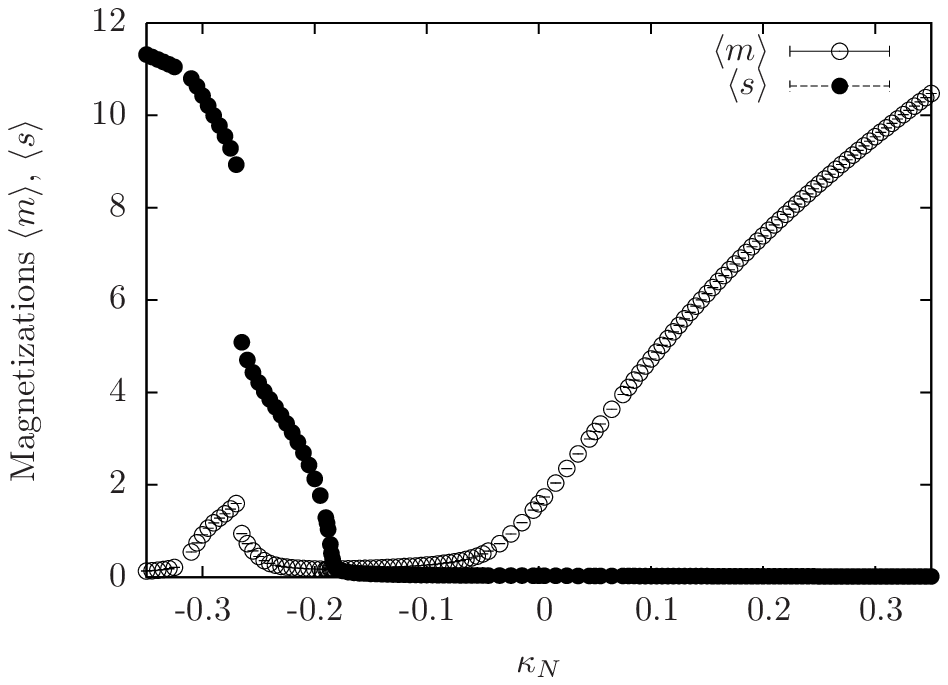}&
\includegraphics[width=0.32\textwidth]{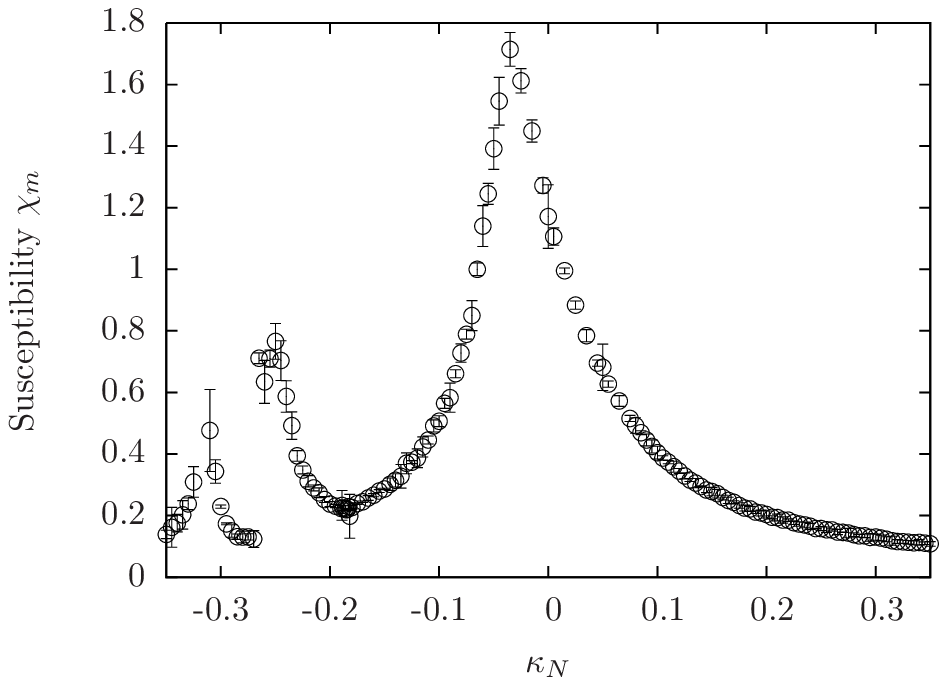}
\\
(a) & (b)  \\
\end{tabular}
\caption{Evidence for the ferrimagnetic phase 
with $\mAvg >0$ and $\sAvg >0$ inside the anti-ferromagnetic phase.
The behaviour of the average magnetization $\mAvg$ and staggered 
magnetization $\sAvg$ is shown in panel (a) as a function of $\kappa_N$ for 
$\tilde y_N=3.162$, $\tilde\lambda_N=0.1$, $N_f=10$, and $L=6$.
The corresponding magnetic susceptibility is shown in panel (b).
From left to right its three observable peaks correspond to the phase transitions
AFM-FI, FI-AFM, and SYM-FM.
From the large $N_f$, $L=\infty$ calculation the ferrimagnetic 
phase was expected to occur approximately at $\kappa_N\le-0.27$.}
\label{fig:ferrimagnetic}
\end{figure}
\ec

Besides these three phases also a fourth, somewhat peculiar phase that can appear at 
intermediate values of the Yukawa coupling constant was predicted by our analytical
investigation. This is the so-called ferrimagnetic (FI) phase where both, the average 
magnetization as well as the average staggered magnetization, are non-zero, 
\ie $\mAvg>0$ and $\sAvg>0$. It was found that such a ferrimagnetic phase should exist deeply inside
the anti-ferromagnetic phase~\cite{Gerhold:2007yb}. In Fig.~\ref{fig:ferrimagnetic} 
we provide evidence for the existence of this ferrimagnetic phase.
Its location within the phase diagram is in good agreement with the analytical
prediction. However, the ferrimagnetic phase is not the prime target for our eventual 
interest and hence we do not further investigate this phase here.

\begin{figure}[htb]
\begin{tabular}{ccc}
\includegraphics[width=0.32\textwidth]{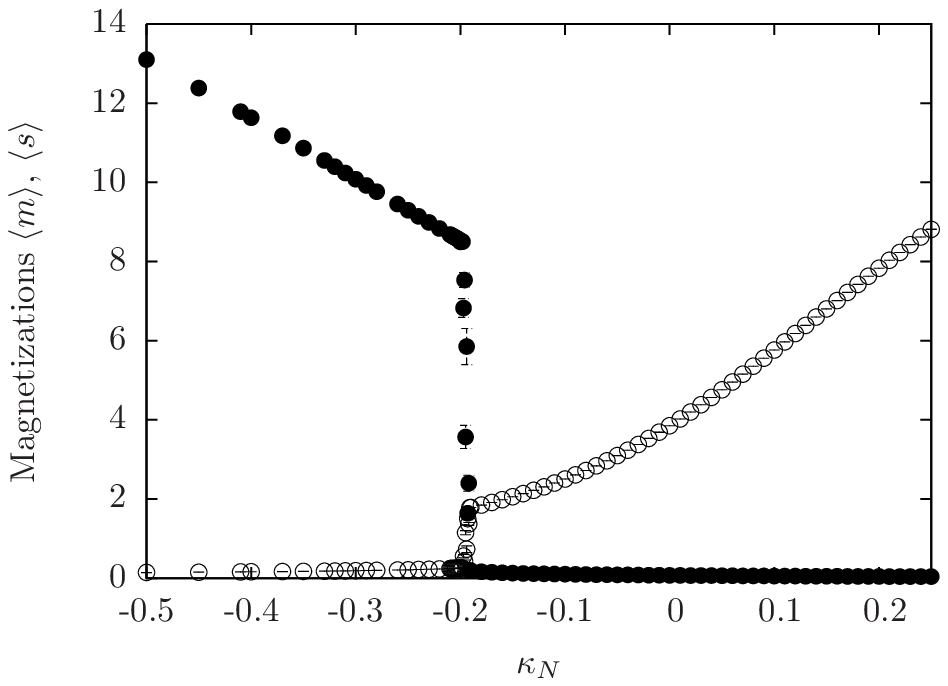}&
\includegraphics[width=0.32\textwidth]{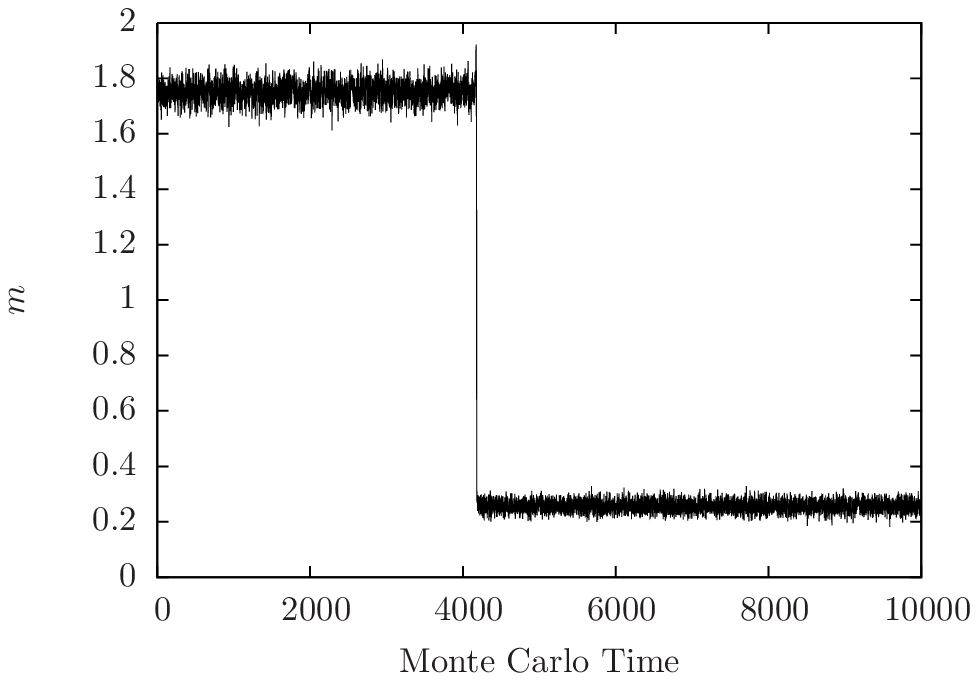}&
\includegraphics[width=0.32\textwidth]{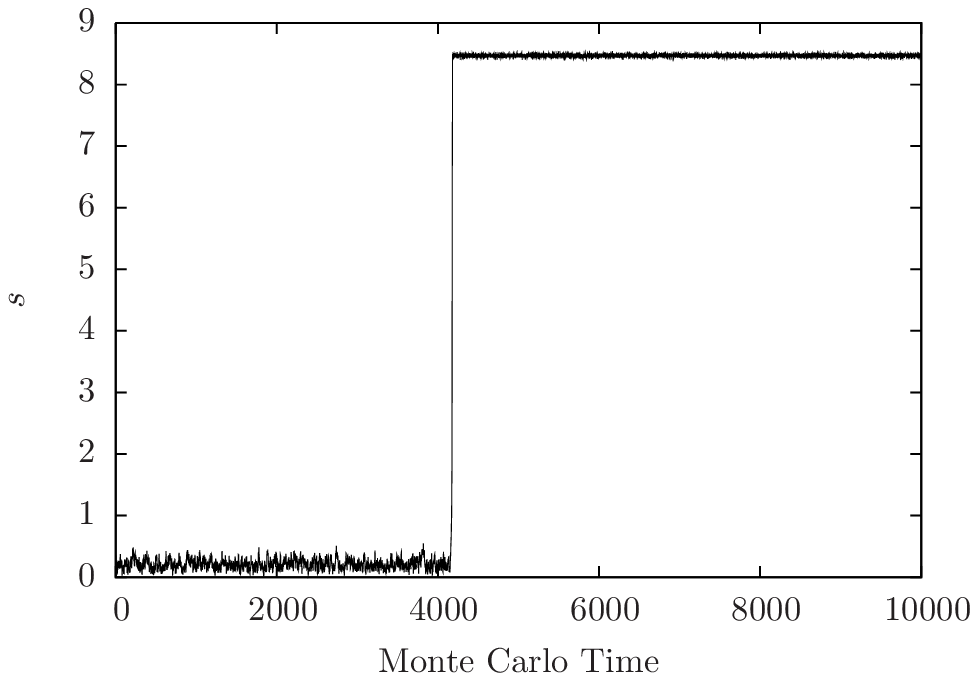}
\\
(a) & (b) & (c)\\
\end{tabular}
\caption{The direct FM-AFM phase transition at intermediate values of the Yukawa
coupling constant. We show, as a function of $\kappa_N$, the behaviour of the 
average magnetization $\mAvg$ and staggered
magnetization $\sAvg$ in panel (a). 
The parameters chosen are $\tilde y_N=6.325$, $\tilde \lambda_N=0.1$, $L=4$, and $N_f=10$.
Panels (b) and (c) show a tunneling event between
two ground states which we take as strong indication for the first order character 
of the phase transition.
The plots show $m$ in panel (b) and $s$ in panel (c), respectively, 
versus the Monte-Carlo time at the hopping parameter $\kappa_N=-0.196$
being very close to its critical value.}
\label{fig:kappascan2}
\end{figure}

Concerning the order of the encountered phase transitions we find that the SYM-FM as well
as the SYM-AFM phase transition seem to be of second order in accordance with the
continuous behaviour of $\mAvg$ and $\sAvg$ as seen e.g. in Fig.~\ref{fig:ferrimagnetic}a.
This is in contrast to the direct FM-AFM phase transition that should occur at intermediate 
values of the Yukawa coupling constant according to our large $N_f$ computation.
From the analytical considerations we expect this transition to be of first order. To clarify 
this we show in Fig.~\ref{fig:kappascan2} an example for such a phase transition as seen in 
the numerical simulations. One can clearly observe an abrupt jump in $\mAvg$ and $\sAvg$ in 
Fig.~\ref{fig:kappascan2}a indicating a discontinuous phase transition. 
In subfigures (b) and (c) we furthermore present an example for a tunneling event between two 
ground states close to the critical value $\kapCrit$ of the hopping parameter serving as another
strong indication for the first order nature of the phase transition at intermediate values
of the Yukawa coupling constant. However, we do not study the order of the phase transition
in great detail here, since this is not in our main interest.

\bc
\ec

Qualitatively, all presented findings are in excellent accordance with our large $N_f$ 
calculations in Ref.~\cite{Gerhold:2007yb}. On a quantitative level, however, the encountered deviations in Fig.~{\ref{fig:phasediagram1}} need to be further addressed. These deviations
can be ascribed to finite volume effects as well as finite $N_f$ corrections.
Here we start with a discussion of the finite volume effects.

\begin{figure}[htb]
\includegraphics[width=5cm]{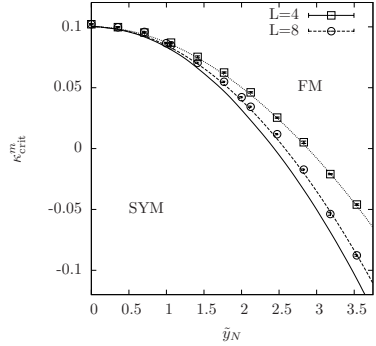}
\caption{A demonstration of finite size effects. We show 
for several selected values of the Yukawa coupling constant
the phase transition points between the ferromagnetic and the symmetric phase as
obtained on a $4^4$-lattice (open squares), and on an $8^4$-lattice (open circles). 
These results are compared to the analytical $L=4$ (dotted), $L=8$ (dashed), and 
$L=\infty$ (solid) phase transition lines determined in the large $N_f$-limit.
The chosen parameters are $\tilde \lambda_N=0.1$ and $N_f = 50$.} 
\label{fig:finitesize1}
\end{figure}

The location of the phase transition points can be strongly altered by
finite size effects. This is illustrated in Fig.~\ref{fig:finitesize1} 
showing some phase transition points from the FM to the SYM phase as obtained 
from our numerical simulations on a $4^4$-lattice (open squares), and on an $8^4$-lattice 
(open circles). One clearly observes that the phase transition line is shifted 
towards smaller values of the hopping parameter when the lattice size is increased. 
This effect can also be anticipated from the analytical computation of the
phase transition line, when one imposes finite lattices also for the minimization 
of the effective potential in the large $N_f$ approximation. Since we want
to demonstrate the finite volume dependence here isolated from the 
$N_f$-dependence, we present the numerical results for the (very large) value
of fermion generations $N_f=50$ and compare them to the analytical $N_f=\infty$
phase transition lines obtained for $L=4$ (dotted line), $L=8$ (dashed line), and 
$L=\infty$ (solid line).
The analytical lines perfectly describe the numerical results and one 
clearly observes the convergence of the numerical results to the analytically 
predicted $L=\infty$ line as the lattice size increases.

However, one remark is in order here for the orientation of the reader, which 
concerns the large $N_f$-computation of the phase transition points in a finite
volume:
The fermionic determinant $det(\fermiMat)$ with $\fermiMat$ given in 
Eq.~(\ref{eq:DefOfFermiMat}) becomes, on a finite lattice, identical to zero 
for completely vanishing Higgs field. On infinite
lattices the zero modes of $\D$ form a set of only zero measure and the integral entering
the effective action can be shown to converge, such that there actually is a symmetric
phase on infinite lattices. For finite $L$ we therefore cannot determine the phase transition
by simply searching that value of the hopping parameter, where the average magnetization vanishes.
Instead we search for that $\kappa_N$, where the minimum of
the effective action $S_{eff}$ becomes flattest, \ie where the second derivative of $S_{eff}$ 
with respect to the magnetization becomes minimal at the location of the minimum. Since the
Higgs field oscillates the stronger around the minimum of the effective action the smaller
its second derivative is, this approach corresponds to finding the phase transition point
by searching for the maximum of the susceptibility.

\bc
\ec

\begin{figure}[htb]
\begin{tabular}{ccc}
\includegraphics[width=0.32\textwidth]{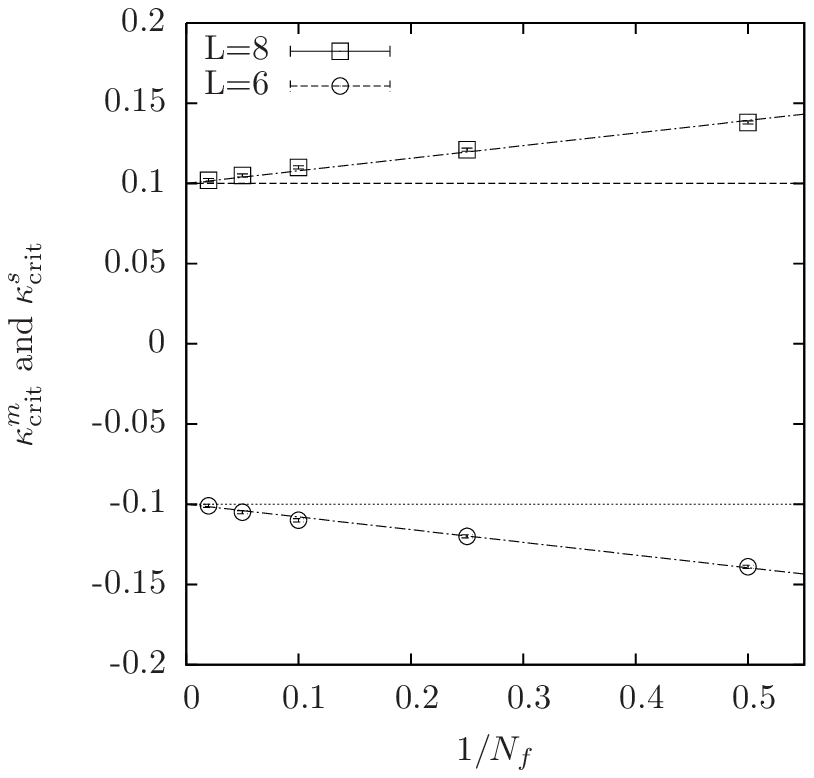}&
\includegraphics[width=0.32\textwidth]{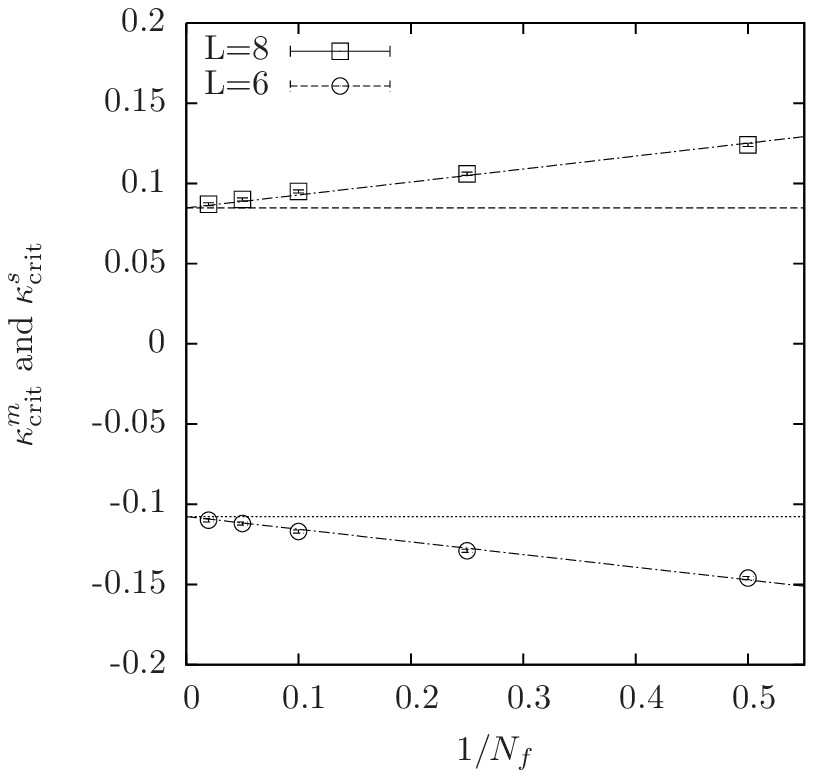}&
\includegraphics[width=0.32\textwidth]{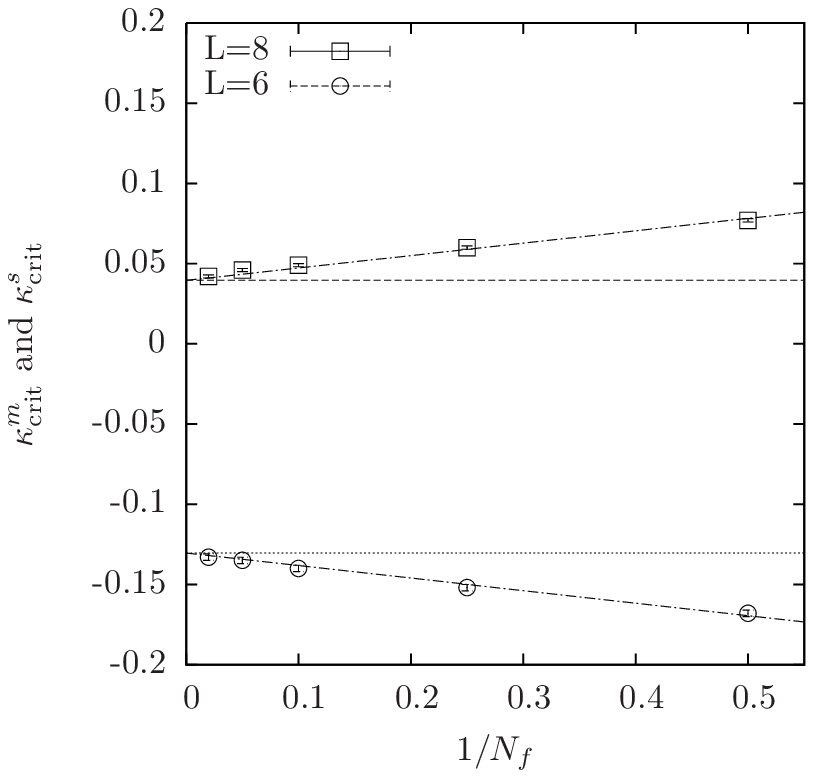}
\\
(a) & (b) & (c) \\
\end{tabular}
\caption{The $N_f$-dependence of the critical hopping parameters
$\kapCrit^m$, $\kapCrit^s$ for the selected Yukawa coupling parameters
$\tilde y_N=0.0$ (a), $\tilde y_N=1.0$ (b), and $\tilde y_N=2.0$ (c).
The data with square symbols were measured on an $8^4$-lattice
while those represented by circles were obtained on $6^4$-lattices.
The analytical, finite volume, large $N_f$ predictions for the SYM-FM (SYM-AFM)
phase transitions are represented by the dashed (dotted) lines. 
The dash-dotted lines are fits of the numerical data to the function 
$f_{m,s}(N_f) = A_{m,s}/N_f + B_{m,s}$ where $B_{m,s}$ is set to the actual analytical prediction
and $A_{m,s}$ is the only free fitting parameter. The results were computed for $\tilde\lambda_N=0.1$.}
\label{fig:NfDependence}
\end{figure}

The $N_f$-dependence of the numerically obtained critical hopping parameters $\kapCrit^m$ and $\kapCrit^s$
is shown in Fig.~\ref{fig:NfDependence} for several selected values of the Yukawa coupling constant. 
One clearly sees that for increasing $N_f$ the numerical results converge very well 
to the analytical finite volume predictions, as expected. It is interesting to note that
the leading term in the finite $N_f$ corrections, \ie the $1/N_f$ contribution, seems to 
be the only relevant correction here, even at the small value $N_f=2$, 
as can be seen in Fig.~\ref{fig:NfDependence} by fitting the deviations to the function 
$f_{m,s}(N_f)=A_{m,s}/N_f$ with $A_{m,s}$ being the only free parameter.
Furthermore, one observes that the critical hopping parameter $\kapCrit^m$ is shifted
towards larger values with decreasing $N_f$ while $\kapCrit^s$ is
shifted towards smaller values.

\bc
\ec

From our findings in this section we finally conclude that the structure of the phase diagram
of the considered Higgs-Yukawa model at small values of the Yukawa coupling constant can be very 
well predicted on a qualitative level by the results of our large $N_f$ analysis. It also gives
a very good understanding of the encountered finite volume effects.

\section{Results for large Yukawa coupling constant}
\label{sec:LargeY}

In this section we want to address the region of large Yukawa coupling constants, \ie $y_N\gg 1$. 
From our large $N_f$ calculations we expect here a ferromagnetic, an anti-ferromagnetic  
and a symmetric phase. The large $N_f$ calculation also revealed that significant finite size 
effects can be present in the symmetric phase which may render its detection difficult. 
This large $N_f$ approach was carried out by scaling the Higgs field and the
coupling constants according to
\beq
\label{eq:LargeNBehaviourOfCouplings2}
y_N = \tilde y_N,\quad  
\lambda_N = \frac{\tilde \lambda_N}{N_f}\,,\quad
\kappa_N = \frac{\tilde \kappa_N}{N_f}\,,\quad
\Phi_n = \sqrt{N_f} \cdot\tilde\Phi_n\,,
\eeq
where the quantities $\tilde y_N$, $\tilde \lambda_N$, $\tilde\kappa_N$,
and $\tilde\Phi_n$ were held constant in the limit $N_f \rightarrow \infty$.

\bc
\begin{figure}[htb]
\begin{tabular}{ccc}
\includegraphics[width=0.32\textwidth]{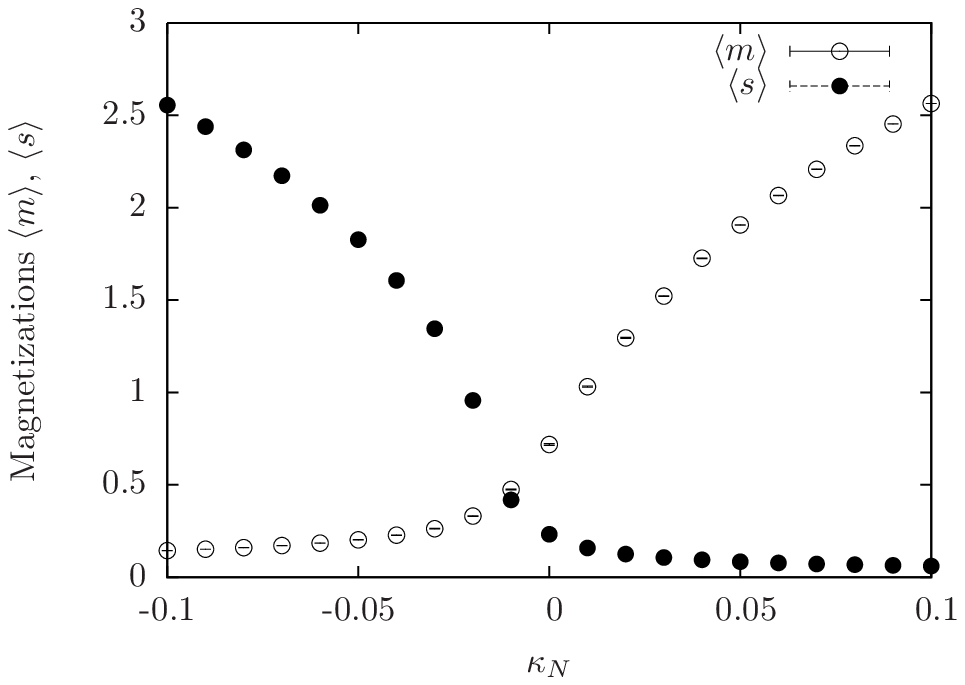}&
\includegraphics[width=0.32\textwidth]{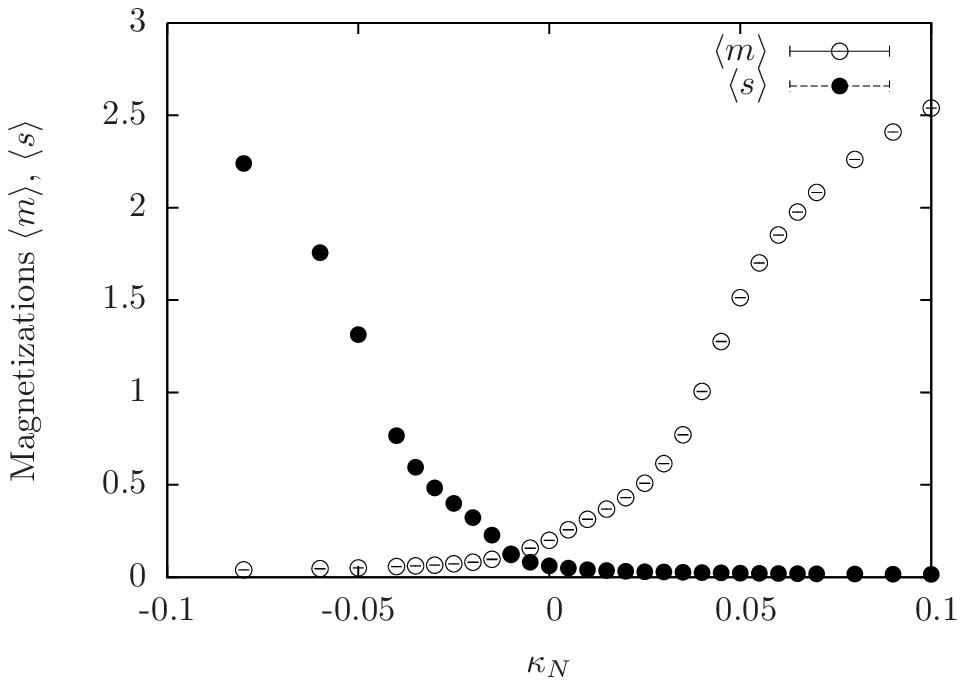}&
\includegraphics[width=0.32\textwidth]{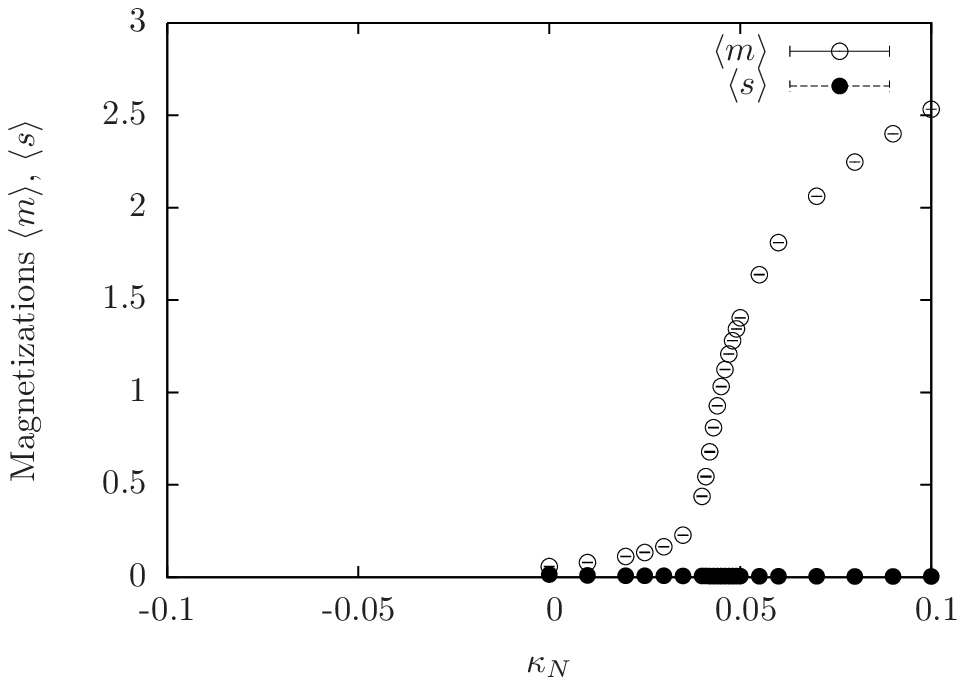}
\\
(a) & (b) & (c) \\
\end{tabular}
\caption{The behaviour of the average magnetization $\mAvg$ and staggered 
magnetization $\sAvg$ as a function of $\kappa_N$ on a $4^4$- (a), $8^4$- (b) 
and $16^4$-lattice (c). In the plots we have chosen 
$\tilde y_N=30$, $\tilde\lambda_N=0.1$ and $N_f=2$.}
\label{fig:kappascan3}
\end{figure}
\ec

\begin{figure}[htb]
\begin{tabular}{cc}
\includegraphics[width=0.32\textwidth]{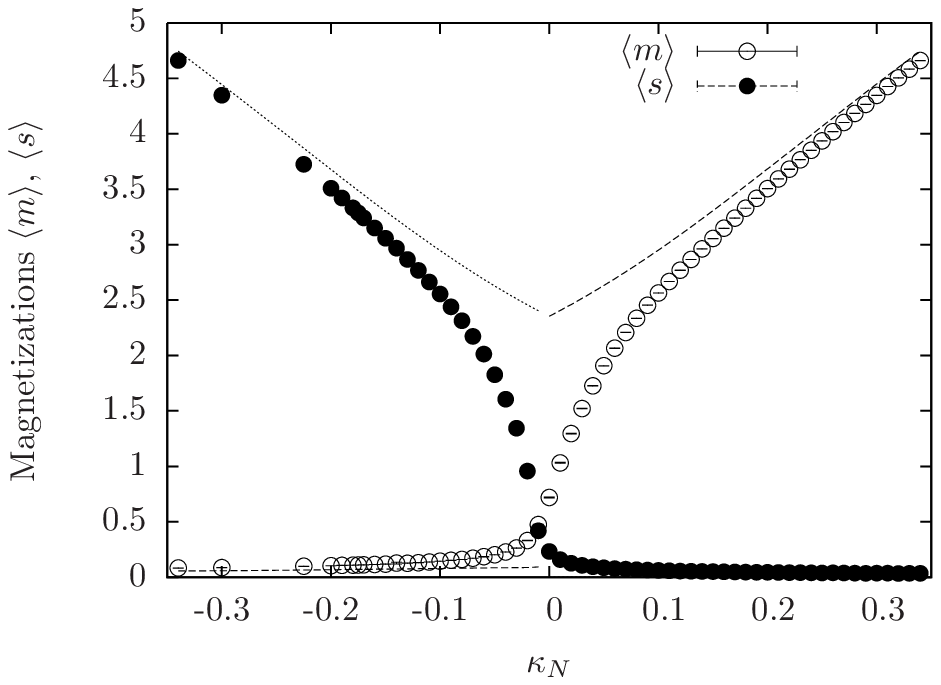}&
\includegraphics[width=0.32\textwidth]{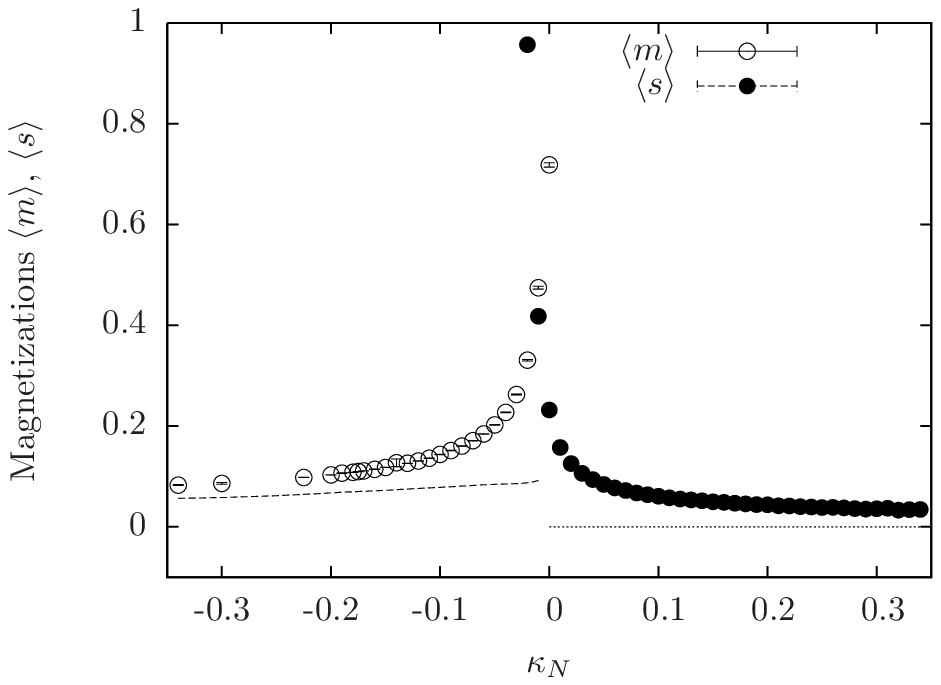}
\\
(a) & (b) \\
\end{tabular}
\caption{Comparison with finite size expectations.
We show again the average magnetizations $\mAvg$ and $\sAvg$ of Fig.~\ref{fig:kappascan3}a
as obtained on the $4^4$-lattice at $\tilde y_N=30$, $\tilde\lambda_N=0.1$, and $N_f=2$.
We compare these magnetizations to the analytical predictions for $\mAvg$ (dashed line)
and $\sAvg$ (dotted line) computed by minimizing the effective, large $y_N$ action of 
Eq.~(\ref{eq:EffectiveActionFullLimitOfLargeY}). 
Panel (b) is just a magnification of plot (a).}
\label{fig:largeYfiniteSE}
\end{figure}

In Fig.~\ref{fig:kappascan3}, we show the numerically obtained values for 
the average magnetizations $\mAvg$ and $\sAvg$ on various sized lattices as 
a function of $\kappa_N$ for a large value of the Yukawa coupling constant
$y_N=30$.
Fig.~\ref{fig:kappascan3} demonstrates that indeed the symmetric phase emerges only on 
sufficiently large lattices, while on small lattices the magnetization 
does not vanish as a function of decreasing $\kappa_N$ even deeply within the anti-ferromagnetic
phase. Instead $\mAvg$ reaches a plateau with a clearly non-vanishing value 
in the limit $\kappa_N\rightarrow -\infty$. This becomes especially well observable
for the smallest considered lattice, the $4^4$-lattice presented in 
Fig.~\ref{fig:kappascan3}a. Thus, one may erroneously conclude that there is no symmetric 
phase at large values of the Yukawa coupling constant, if one considers too small
lattices. However, the plateau value of $\mAvg$ is fully consistent with our analytical results 
predicting a finite volume effect causing a non-vanishing magnetization $\mAvg>0$ also for 
arbitrarily negative values of $\kappa_N$.
To demonstrate this latter statement we restate here one result of 
Ref.~\cite{Gerhold:2007yb} for the effective action of a field configuration 
in terms of its magnetizations $m$ and $s$ in the large $y_N$-limit, reading
\beq
\label{eq:EffectiveActionFullLimitOfLargeY}
S_{eff}[\Phi] = S_\Phi -N_f \cdot \sum\limits_{n} 8\log\left|m+s\cdot(-1)^{\sum\limits_\mu n_\mu}\right|
- N_f \cdot 8\log\left|\tilde m \right| 
- N_f \cdot 56\log\left| \tilde m^2 -\tilde s^2\right|
\eeq
with the abbreviations
\beq
\tilde m=\frac{m}{m^2-s^2} \quad \mbox{and}\quad
\tilde s=\frac{s}{s^2-m^2}.
\eeq
Considering only the ground state of this effective action one cannot correctly
predict the phase transition of the model, as discussed in Ref.~\cite{Gerhold:2007yb}.
However, it is sufficient to correctly predict the behaviour of $\mAvg$ and $\sAvg$ 
in the limit of large negative (and positive) values of the hopping parameter $\kappa_N$,
as demonstrated in Fig.~\ref{fig:largeYfiniteSE}, where we plot again
the average magnetizations for the $4^4$-lattice together with the finite volume 
analytical expectations, obtained by minimizing the effective action of 
Eq.~(\ref{eq:EffectiveActionFullLimitOfLargeY}). The convergence 
of the numerical results to the analytical finite volume prediction is very well observed
in Fig.~\ref{fig:largeYfiniteSE}b. 

\bc
\ec

We remark that the non-vanishing plateau as well
as the asymmetry in $\mAvg$ and $\sAvg$ are both caused by the term $\log\left|\tilde m \right|$
appearing in Eq.~(\ref{eq:EffectiveActionFullLimitOfLargeY}). This term as well as the very last
one in this equation do, however, not scale proportional to the volume $L^4$ in contrast to all 
other contributions to the effective action.
Its influence therefore eventually disappears as the lattice size increases. This is exactly what
is observed here.

\bc
\begin{figure}[htb]
\begin{tabular}{ccc}
\includegraphics[width=0.32\textwidth]{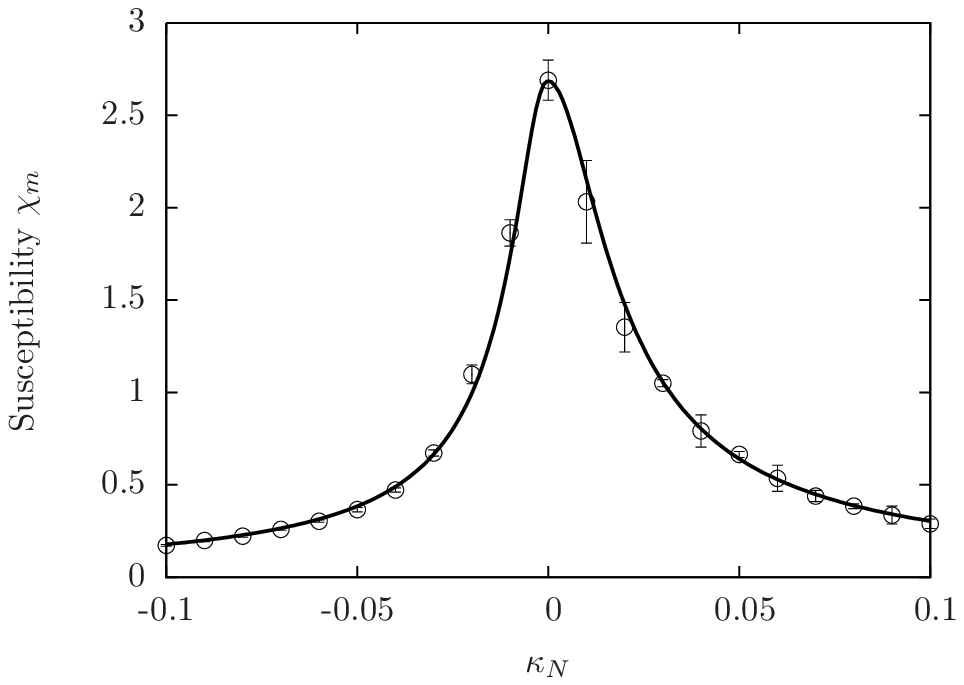}&
\includegraphics[width=0.32\textwidth]{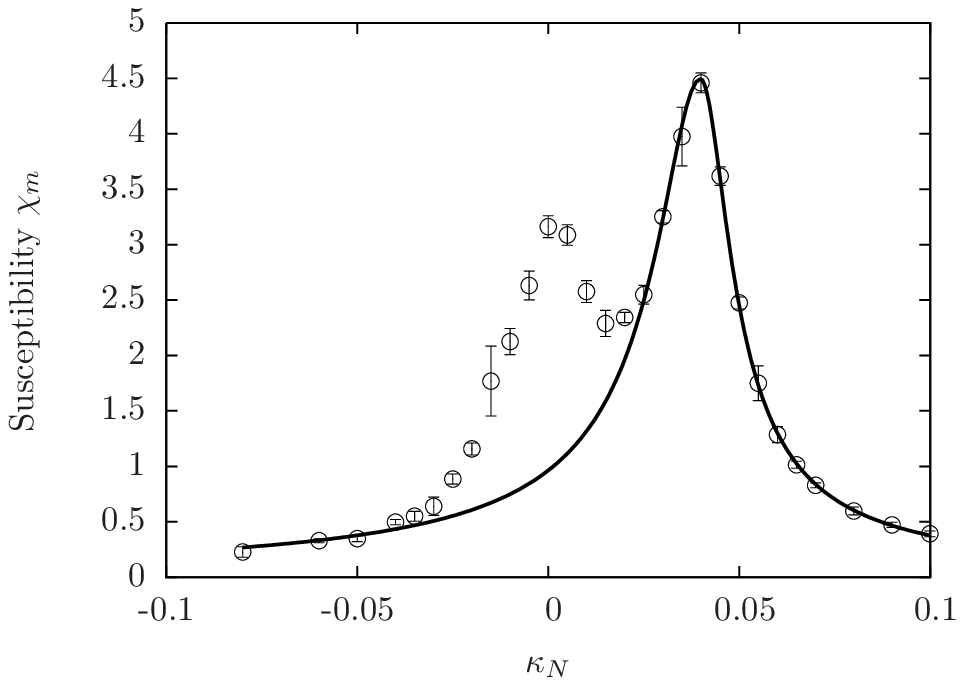}&
\includegraphics[width=0.32\textwidth]{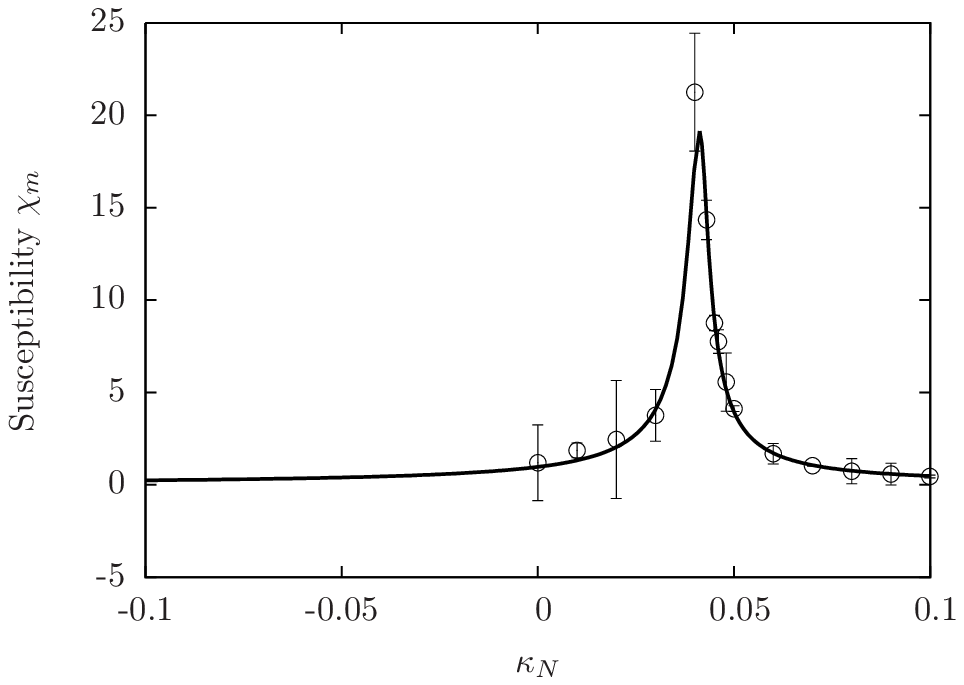}
\\
(a) & (b) & (c) \\
\end{tabular}
\caption{The behaviour of the magnetic susceptibility $\chi_m$
as a function of the hopping parameter $\kappa_N$ on a $4^4$- (a), $8^4$- (b) 
and $16^4$-lattice (c). In the plots we have chosen 
$\tilde y_N=30$, $\tilde\lambda_N=0.1$ and $N_f=2$.
The fit in panel (b) is only applied to those points with $\kappa_N\ge0.025$
or $\kappa_N\le-0.05$ in order to reduce the influence of the unphysical peak
at $\kappa_N=0.0$. Note the changing scale in the three plots.}
\label{fig:kappascan4}
\end{figure}
\ec

In Fig.~\ref{fig:kappascan4} we show the susceptibilities $\chi_m$ corresponding 
to the magnetizations in Fig.~\ref{fig:kappascan3}.
For the smallest lattice, \ie the $4^4$-lattice, one observes only one peak
in the magnetic susceptibility, centered at $\kappa_N=0$. From this result
one could conclude that the phase transition point is located at $\kappa_N=0$,
excluding a symmetric phase, since the staggered susceptibility reaches its
maximum at the same value of $\kappa_N$. However, with increasing lattice sizes
a second peak develops in the susceptibilities. This is very well observed in
Fig.~\ref{fig:largeYfiniteSE}b corresponding to the larger $8^4$-lattice. It shows
that indeed two distinct  peaks emerge on this intermediate lattice. 
It is actually this second peak, centered around $\kappa_N=0.04$ in this case, 
that correctly describes the {\it physical} phase transition between the ferromagnetic 
and the symmetric phase, while the first one is only caused by the finite volume terms
discussed in~\cite{Gerhold:2007yb}, which do not scale with the lattice volume.
Its height is therefore at most constant in contrast to the physical peak, which
grows with increasing lattice volume. On the largest presented lattice, the 
$16^4$-lattice, the physical peak at $\kappa_N=0.04$ completely dominates the
scene and the former small volume peak at $\kappa_N=0$ has disappeared, presumably
hidden beneath the large error bars at $\kappa_N=0$. 

\bc
\begin{figure}[htb]
\includegraphics[width=5cm]{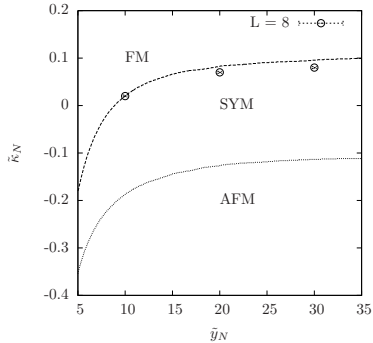}
\caption{The phase diagram for large values of the Yukawa coupling constant. 
We show the numerical results for the phase transition points to the ferromagnetic phase
as obtained on an $8^4$-lattice at $\tilde\lambda_N=0.1$, and $N_f=2$.
We could not reliably determine the phase transition points to the anti-ferromagnetic
phase, as explained in the main text. 
We compare these results to the prediction of our large $N_f$ calculation 
for the SYM-FM transition (dashed line). The SYM-AFM phase transition is marked
by the dotted line.} 
\label{fig:phasediagram3}
\end{figure}
\ec

We have then determined the SYM-FM phase transition points by fitting the physical peaks
in the magnetic susceptibility $\chi_m$ on the intermediate $8^4$-lattices to the finite 
volume expectation in Eq.~(\ref{finitesizesus}) by taking only the points belonging to the 
physical peak into account as demonstrated in Fig.~\ref{fig:kappascan4}b. 
We remark here that we do not provide any data for the SYM-AFM phase transition
because the phase transition is not reliably detectable on the $6^4$-lattice, due to the
finite volume effects discussed above, and the $8^4$-simulations are not practicable in
the anti-ferromagnetic phase with our algorithm as explained in Section~\ref{sec:SmallY}.

In Fig.~\ref{fig:phasediagram3} we finally summarize the obtained phase transition points 
together with the analytical $N_f=\infty$, $L=\infty$ expectation of the phase structure 
at large values of the Yukawa coupling constant. 
Qualitatively, the picture we obtain from the numerical simulations 
is in full accordance with the results from the large $N_f$ approximation: there 
are second order phase transitions separating a ferromagnetic phase from a symmetric 
phase. In this symmetric phase strong finite size effects are
encountered such that only for large lattice sizes this symmetric phase can be 
identified. 
Quantitatively, the numerical results deviate from the analytical expectation due to the 
finite settings $N_f=2$ and $L=8$, but are still in good agreement.

In this section we found that a symmetric phase at large values of the
Yukawa coupling constant does indeed exist although its existence is obscured on too small
lattices by strong finite size effects, and that its location within the phase diagram is in good 
agreement with the analytical large $N_f$ predictions. 
We remark here that the existence of a symmetric phase at strong Yukawa couplings has also
been observed and discussed in Ref.~\cite{Giedt:2007qg}.
From our findings we thus conclude that the analytical large $N_f$ calculations describe 
the phase structure of the considered Higgs-Yukawa
model at large values of the Yukawa coupling constant very well.

\section{Summary and outlook}

In this paper we have studied by numerical simulations the phase structure of a chirally invariant
lattice Higgs-Yukawa model, originally proposed by L\"uscher, in order to check
the validity of our earlier analytical investigation of its phase structure. These earlier
calculations have been performed in the large $N_f$-limit for small and for large values
of the Yukawa coupling constant. 

In Section~\ref{sec:SmallY} we compared the numerical to the analytical results at 
small values of the Yukawa coupling constant. We started with a discussion of the
qualitative structure of the phase diagram. For that purpose we presented our numerical 
results for the phase transition lines obtained at $N_f=10$ on some $8^4$- and $6^4$-lattices
and compared them to the analytically computed $N_f=\infty$, $L=\infty$ predictions.
Qualitatively, the numerical and analytical results are in very good agreement: As expected we 
clearly observe a symmetric (SYM), a ferromagnetic (FM) and an anti-ferromagnetic (AFM) phase.
With increasing Yukawa coupling constant the symmetric phase strongly bends downwards to smaller 
values of the hopping parameter. In particular, we find that the obtained phase structure resembles 
that of earlier Higgs-Yukawa models on a qualitative level. One peculiarity
in these types of models is the emergence of a ferrimagnetic (FI) phase
with $\mAvg>0$ and $\sAvg>0$ deeply inside the anti-ferromagnetic phase. This phase was
also predicted by the analytical investigation of the model and it is located at the predicted
position within the phase diagram. 
Furthermore, the data also support very well the analytical expectations concerning the order
of the encountered phase transitions. The SYM-FM and the SYM-AFM phase transitions were supposed
to be of second order while the direct FM-AFM transition was predicted to be of first order.
Although we did not study that in great detail, since this aspect was not in out main interest, the
obtained lattice results are in very good agreement with these analytical findings.

We then turned towards the quantitative discussion of the encountered deviations between
the numerical finite volume, finite $N_f$ results and the presented analytical $N_f=\infty$,
$L=\infty$ calculations. Firstly, we showed that finite volume effects alter the location of 
the phase transition lines strongly. In order to isolate the finite volume effects from
the $1/N_f$ corrections we presented numerical results for the phase transition points for 
the choice of the very large number of fermion generations $N_f=50$. In that setting we could show that
the finite volume effects are in excellent agreement with the analytical finite volume
predictions. We then demonstrated the strength of the $1/N_f$ corrections by presenting
the numerically obtained phase transition points at smaller values of $N_f$ for some
selected Yukawa coupling parameters. We found that the $1/N_f$ corrections drive the critical
hopping parameters towards larger values for the case of the SYM-FM phase
transition and towards smaller values for the SYM-AFM transition. Besides that these 
corrections do not change the qualitative phase structure of the model. 

We then discussed the phase structure at large values of the Yukawa coupling constant in
Section~\ref{sec:LargeY}. In particular we showed that there is actually a symmetric phase in this
regime of the Yukawa coupling constant and that it is located at the expected position within the
phase diagram. We also demonstrated that this symmetric phase becomes unobservable on too small
lattices due to strong finite volume effects, as derived in our earlier studies, preventing the 
Higgs field expectation value from vanishing. We furthermore showed that the behaviour of the 
magnetization at large negative (and positive) values of the hopping parameter $\kappa_N$ can be 
very well described by taking these finite volume contributions into account. The emergence of the 
symmetric phase with increasing lattice size could also clearly be observed in the presented plots 
of the magnetic susceptibility $\chi_m$ (Fig.~\ref{fig:kappascan4}).
Finally, we presented our numerical results for the critical hopping parameters $\kapCrit^m$ of 
the SYM-FM phase transition at large Yukawa coupling constants. We compared them to the analytical 
large $N_f$ predictions and found them to be in good agreement even though the numerical 
simulations were performed at $N_f=2$.

We end with a short outlook about our next steps concerning the further investigation of the
presented Higgs-Yukawa model: We have started the implementation of a PHMC algorithm~\cite{Frezzotti:1998eu} with
which the simulation will become possible at arbitrary values of $N_f$, in particular at
the physically interesting setting $N_f=1$. Having the qualitative phase diagram of the model
at hand we will then search for the physical region of the parameter space, reproducing the 
top quark mass, eventually allowing to find upper and lower bounds for the Higgs boson mass.

\section*{Acknowledgments}
We thank the "Deutsche Telekom Stiftung" for supporting this study by providing a Ph.D. scholarship for
P.G. We further acknowledge the support of the DFG through the DFG-project {\it Mu932/4-1}.
We are grateful to Joel Giedt, Julius Kuti, Michael M\"uller-Preussker, Erich Poppitz, and Christopher Schroeder
for enlightening discussions and comments. In particular we want to 
express our gratitude to Julius Kuti for inviting P.G. to his group at the University of
California, San Diego.

\bibliographystyle{unsrtOWN}
\bibliography{HiggsYukawaPhaseDiagramSimulation}

\end{document}